\documentclass[12pt,a4paper]{article}
\usepackage{graphicx}
\input{epsf}
\usepackage{amsmath}
\usepackage{amssymb}
\usepackage{caption}
\usepackage{txfonts}
\usepackage{url}
\usepackage{mcite}
\usepackage{textpos}
\begin{document}

\title{Physics highlights at ILC and CLIC} 

\author{S. Luki\'c\thanks{E-mail:slukic@vinca.rs}\\
{\small Vin\v{c}a Institute of Nuclear Sciences, University of Belgrade, Serbia}\\
On behalf of the FCAL Collaboration and the \\CLIC detector and physics study \footnote{Presented at the 12th International School-Seminar \textit{The Actual Problems of Microworld Physics} in Gomel, Belarus, July 22 - August 2, 2013} \footnote{This work was carried out in the framework of the CLICdp collaboration}
}

\date{}
\maketitle

\begin{textblock}{1}[1,0](1,-8)
\begin{flushright}
\textbf{CLICdp-Conf-2013-001} \\
October 28, 2013
\end{flushright}
\end{textblock}

\begin{abstract}
In this lecture, the physics potential for the $e^+e^-$ linear collider experiments ILC and CLIC is reviewed. The experimental conditions are compared to those at hadron colliders and their intrinsic value for precision experiments, complementary to the hadron colliders, is discussed. The detector concepts for ILC and CLIC are outlined in their most important aspects related to the precision physics. Highlights from the physics program and from the benchmark studies are given. It is shown that linear colliders are a promising tool, complementing the LHC in essential ways to test the Standard Model and to search for new physics. 
\end{abstract}

\section{Introduction}

In the European strategy for particle physics, linear electron-positron colliders represent an important component of the future High-Energy physics program. They are designed for precision measurements, complementary to the present Large Hadron Collider (LHC), as well as to its possible upgrades and successors at CERN \cite{EuStrat13}. At present, two international projects are devoted to the design of the future linear colliders - the International Linear Collider (ILC) \cite{ILCTDR2} and the Compact Linear Collider (CLIC) \cite{CLICCDR12acc}. 

One of the main goals of the linear collider experiments is to test the Standard Model (SM), in particular regarding the mechanism of the Electroweak Symmetry Breaking (EWSB). The recent discovery of a new scalar boson at LHC \cite{ATLAS12, *CMS12}, with properties consistent with those of the SM Higgs boson, has given a very strong impetus to this area of research. Precision measurements of the Higgs sector can test different existing theories describing EWSB. Another important area is the search for new physics. This area is driven by the quest to resolve open questions in particle physics, as well as by the evidence from cosmology of phenomena that cannot be explained within the framework of the SM.

The crucial motivation and potential of the linear colliders is that of fundamental advance in knowledge. This lecture will underline \emph{new} knowledge that can be gained through precision measurements \cite{ILCTDR2, CLICCDR12PD}.

\subsection{The experimental environment at hadron versus lepton colliders}

The difference in nature of the colliding particles lies at the origin of all of the major differences between the hadron and lepton collider experiments. 

Since hadrons are compound objects, the initial state of individual partons is not uniquely defined. In the general case, initial states are realized as quantum superposition of states distributed according to the proton structure functions. In the analysis, distributions of initial parton states are calculated using QCD models tuned to data from deep inelastic scattering experiments \cite{HERA10}. 

At lepton colliders the colliding particles are elementary, therefore the inital state is well defined at the fundamental level. This allows for full reconstruction of the final state from conservation principles, up to the distribution of initial center-of-mass (CM) energies. The distribution of initial particle energies due to beam-beam effects can be precisely measured in the course of the experiment \cite{Moe00, Rim07, Luk13, Boz13, Sailer13}. 

Each collision at a hadron collider creates a large number of elementary processes. Most of these processes represent background for the physics analysis, and deposit high doses of radiation energy in the detector. Complex trigger schemes, with the retention rate of only $\sim 1$ event in $10^6$, have to be employed during the data taking in order to select events that are of interest for the physics analysis. Moreover, due to high radiation levels, an important issue for the detector design is the radiation hardness of detectors at all angles.

By contrast, the total cross section at lepton colliders is relatively small. The total radiation levels are moderate, and the radiation dose does not represent an issue for the detector design except in the very forward region. The pulsed beam structure allows for the readout of all detector data. The readout is thus triggerless, and the experiment is cleaner with regards to the physics background. In terms of cross sections, lepton colliders have high sensitivity to electroweak processes, allowing very precise measurements in the Higgs sector, as well as in the search for new physics.

\section{The accelerator concepts}

\subsection{ILC accelerator}

The electrons for the ILC beam are produced by a polarized photocathode DC gun electron source. The electrons are first accelerated to 15 GeV in the bunch compressor, and then in the main linac to the nominal energy. The positrons are generated by pair conversion of high-energy photons produced by passing the high-energy electron beam through an undulator. The beam acceleration in the ILC main linac is provided by niobium superconducting nine-cell cavities. The beam delivery systems then bring the two beams into collision with a crossing angle of 14 mrad. 

The ILC beam is structured in bunch trains arriving at a rate of 5 Hz. The length of the bunch trains is 1 ms. The bunch spacing within the train is 370~ns, allowing full separation of events from different bunches by detector timing techniques. At 500 GeV in CM, each bunch contains $2 \times 10^{10}$ electrons in a quasi-Gaussian spatial distribution with $\sigma_x$ = 470 nm, $\sigma_y$ = 5.9 nm and $\sigma_z$ = 300 $\muup$m, resulting in instantaneous luminosity of $2 \times 10^{34} cm^{-2} s^{-1}$ \cite{ILCTDR3}.

The present state of the art of the superconducting RF technology is a result of several decades of development \cite{TESLA}. The field gradient in superconductors is limited by the field emission, as well as by the quench-causing surface defects. The FLASH FEL facility at DESY, Hamburg, has been in operation since 2004 with an average gradient of 20 MV/m in the main accelerator \cite{FlashFEL}. For the European XFEL program, gradients up to 35 MV/m have been realised in TESLA prototype cryomodules using surface electropolishing \cite{Lil04}. Many of the beam-tuning techniques required by the ILC have also been demonstrated at the FLASH FEL. R\&D on creation of small emittance beams, as well as their focusing and alignment, is done at the Accelerator Test Facility (ATF) at KEK, Japan. Suppression of the electron cloud formation in the beam tube is studied within the CesrTA program at the Cornell University \cite{ILCTDR3}.

\subsection{CLIC accelerator}

The main objective of the CLIC project is to build a linear collider for the multi-TeV range at reasonable cost and size. This requires very high acceleration gradients, which cannot be reached with the superconducting technology. Therefore, CLIC is based on the novel two-beam acceleration technology, in which a low-energy high-current drive beam provides the RF power for the acceleration of the physics beam. The acceleration cavities for the main beam operate at room temperature, and can sustain field gradients over 100 MV/m .

In order to maintain a low breakdown rate, the length of the CLIC bunch train has to be limited to about 150 ns. At the same time, in order to achieve high luminosity, the bunch focusing has to be very strong, the bunch population high, and the bunch spacing very short. In the standard beam parameter set at 3 TeV, RMS bunch dimensions are $\sigma_x$ = 40 nm, $\sigma_y$ = 1 nm and $\sigma_z$ = 44 $\muup$m, bunch population is $3.7 \times 10^9$ and bunch spacing is only 0.5 ns, which results in a luminosity value of $5.9 \times 10^{34} cm^{-2} s^{-1}$ \cite{CLICCDR12acc}

The two-beam acceleration scheme is the subject of study of the CTF3 project at CERN. Some of the most important milestones achieved until now include the generation of an acceleration field well above 100 MV/m, as well as excellent performance of the accelerating structures at the nominal field of 100 MV/m without beam load \cite{CLICCDR12acc}.

\section{The detectors for a linear collider}

\begin{figure}
\centering
\includegraphics[width=0.8\textwidth]{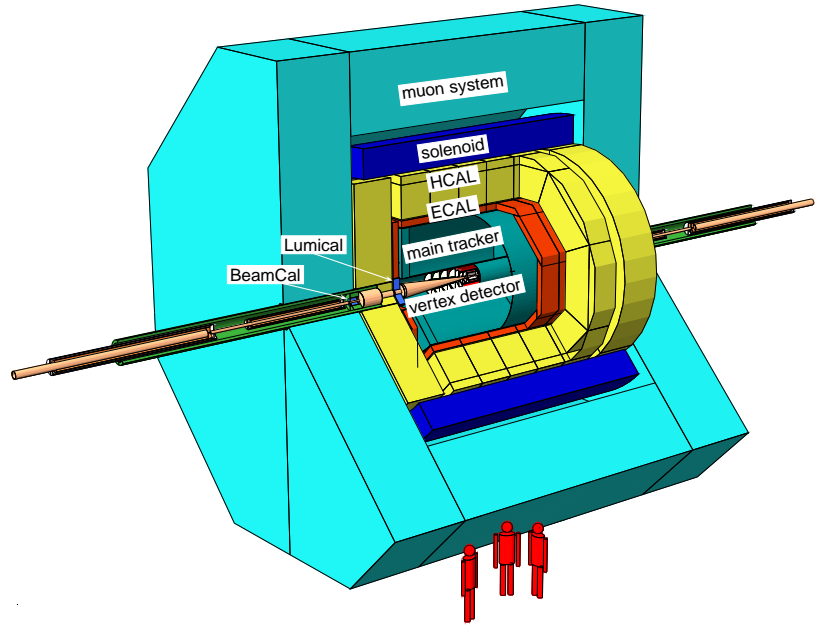}
\caption{\label{fig-detector}Schematic view of the International Large Detector concept for the linear collider.}
\end{figure}

Two detector concepts are foreseen for the future linear collider, the International Large Detector (ILD) and the Silicon Detector (SiD) \cite{ILDLoI, SiDLoI}. The basic layout of both detectors is very similar (Fig. \ref{fig-detector}). The main tracker of ILD is based on a Time-Projection Chamber (TPC) for quasi-continuous track reconstruction, supplemented by inner and outer barrel silicon strip layers for precise track reference, and a forward silicon strip tracker. SiD is a compact cost-effective detector with a 5 Tesla magnetic field and all-silicon tracking with 5 layers in the barrel and 7 layers in the endcap region. Both detectors are planned to be implemented using a push-pull configuration which allows installing one detector in the beam line while the other is in the hangar for maintenance.

\begin{description}
\item[Vertex detector] consists of a number of thin pixelized semiconductor layers with extremely light support structure. Its purpose is to allow reconstruction of the secondary vertices by precise tracking, avoiding multiple scattering in the material. The innermost barrel layer has a radius of 16 mm. Depending on the angle and energy of the detected particles, the impact parameter resolution is between a few $\muup$m and several ten $\muup$m \cite{ILCTDR2}.

\item[Main tracker] performs precise 3D reconstruction of particle tracks in the magnetic field. Both the ILD and the SiD tracker systems satisfy the design goal for the transverse momentum resolution of charged particles of $\Delta(1/p_T) \le 2 \times 10^{-5} \text{GeV}^{-1}$ \cite{ILDLoI, SiDLoI}.

\item [Electromagnetic calorimeter] (ECAL) reconstructs electromagnetic (EM) showers and provides distinction of EM showers from the hadronic ones. The ECAL is designed with tungsten absorber layers, interspersed with scintillator tiles or silicon pads with high granularity. Because of the large difference between EM radiation length and nuclear interaction length in tungsten, hadronic showers develop slower, and start at larger depth of material than the EM showers. 

\item[Hadronic calorimeter] (HCAL) is designed with steel absorbers and either scintillator tiles or gas detectors, with sufficient thickness for full containment of hadronic showers. The main aim of HCAL is to measure the energy of neutral hadrons, identified by the absence of tracks in the main tracker. The long EM radiation length in steel allows fine longitudinal sampling with a reasonable number of layers. 

\item[Forward calorimeters] In the very forward region of the detector, two calorimeters are installed, LumiCal for precise luminosity measurement by counting Bhabha-scattering events and BeamCal for fast luminosity estimate and for monitoring of the beam parameter by measurement of beam-induced processes at low angles. Both calorimeters are centered around the outgoing beam axis. BeamCal covers angles from below $1^\circ$ to $2^\circ$, and LumiCal from about $2^\circ$ to $6^\circ$. LumiCal is designed for precise reconstruction of EM showers, while the main challenge for BeamCal is radiation hardness because of the relatively high radiation dose at small angles \cite{Abr10}. 

\end{description}

Both detector concepts have also been adapted for the CLIC environment \cite{CLICCDR12PD}. Main differences include calorimeter thickness, and the use of tungsten absorber, in order to contain higher-energy showers, higher semiconductor granularity to cope with the occupancy and a larger diameter of the innermost barrel layer of the vertex detector because of higher radiation. 

\subsection{Particle flow calorimetry}

\begin{figure}
\centering
\includegraphics{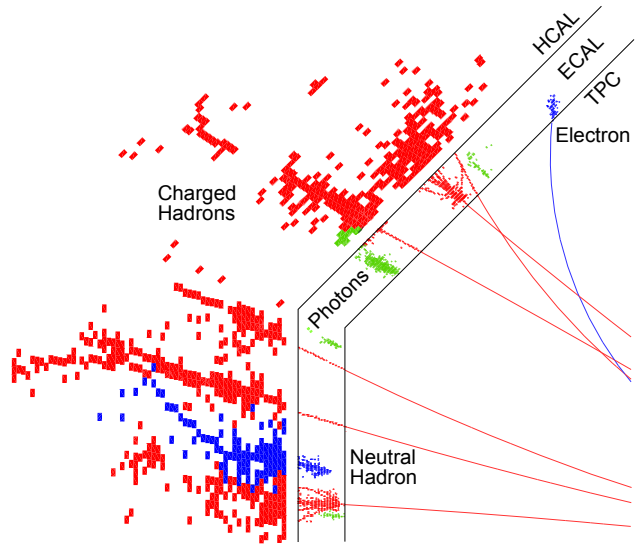}
\caption{\label{fig-PF}Particle tracks from a simulated jet in CLIC\_ILD (from Ref. \protect\cite{Mar12}).}
\end{figure}

About 10\% of energy of a typical jet is carried by long-lived neutral hadrons, 62\% by charged particles, mostly hadrons, another 27\% by photons and 1.5\% by neutrinos \cite{Mar12}. If visible jet energy is entirely reconstructed from calorimetric information, the precision is limited by the relatively poor energy resolution of HCAL. To improve the jet energy resolution, the Particle Flow concept aims at full identification of all constituent particles in the detector system, so that charged particle energies can be reconstructed from track curvatures. To this aim, finely granulated calorimeters are required to separate and reconstruct showers. This allows for precise reconstruction of invariant masses of jets and accurate identification of physics events. 

Figure \ref{fig-PF} shows a typical reconstructed jet in a simulation of the CLIC\_ILD. Electrons are identified by a curved track in the main tracker, and a fast-developing shower in the ECAL. Showers induced by photons develop fast as well, but there is no associated track in the tracker, due to the low interaction cross section in the low-density material of the tracker. Hadrons develop showers slower and deposit a large fraction of their energy in the HCAL. Neutral hadrons have no associated track in the tracker. It has been shown by simulation that a jet energy resolution between 3 and 3.7\% is achievable in the entire energy range from 0 to 1.5 TeV in the barrel region of the CLIC\_ILD \cite{Mar12}.

\section{Physics program}

Both in the ILC and the CLIC projects, the accelerator is planned to be built in stages defined with the physics potential in view. At each stage, each of the accelerators can be tuned to lower energy, at some cost in luminosity. 

\begin{figure}
\centering
\includegraphics{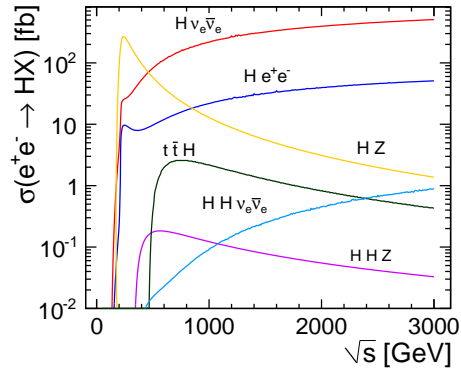}
\caption{\label{fig-higgs-xs}Higgs production cross section as a function of the CM energy.}
\end{figure}

At 250 GeV, the Higgs boson production by Higgsstrahlung (HZ) has its maximum (see Fig. \ref{fig-higgs-xs}). This point gives access to first precise measurements of the Higgs couplings and mass. At an accelerator built for a 250 GeV CM energy, high-precision $W$ mass study can be also performed by tuning the accelerator down to 160 GeV in CM. The "Giga-Z" program \cite{Haw00} is also within reach at 91 GeV in CM, provided that luminosity can be measured with the required precision. The 250 GeV CM energy is the first stage of the ILC program.

At 350 GeV, the Higgsstrahlung and the $WW$ fusion ($H\nu_e\bar\nu_e$) processes of Higgs production have comparable cross sections. This allows for the measurement of absolute Higgs couplings, as well as a model-independent measurement of the total Higgs width. The top quark mass can be precisely measured in a production threshold scan. In the CLIC project, 350 GeV is considered as the first energy stage. In the ILC program, the top-pair threshold scan is performed at the 500 GeV stage with the accelerator tuned down to 350 GeV.

Above 350 GeV, Higgs production is accessed predominantly by the $WW$ fusion, allowing higher precision of Higgs couplings. Precise measurement of most couplings of the gauge bosons is best performed at 500 GeV. New physics is best accessed at higher energies. Production thresholds for supersymmetric particles are expected to start just below 1 TeV, and the mass reach for searches such as the search for the $Z'$ boson in the $e^+e^- \rightarrow f \bar{f}$ channel is higher for higher CM energies.

The ILC program is thus planned in three building stages. The 250 GeV stage for the first precise measurements of the Higgs sector, the nominal design energy of 500 GeV, and the ultimate CM energy of 1 TeV, achievable by extension of the main linac and the use of cavities with higher gradient, so that the total length of the facility reaches 50 km.

The CLIC machine is designed for searches for new physics at multi-TeV energies, with the goal to reach 3 TeV in the CM frame. Currently a lowest energy stage of 350 GeV is considered, followed by an upgrade to 1.4 TeV, and the final stage of 3 TeV, for which the accelerator facility will be 48 km long.

Benchmark studies of physics performance of the ILC and CLIC experiments have been performed using dedicated tool-chains consisting of process generation with realistic beam- and Beamstrahlung spectra, relevant physics- and beam-induced background, complete and realistic simulation of the interaction of the final particles with the detector, as well as event reconstruction using Particle Flow algorithms developed for the linear colliders.

\subsection{Highlights from the Standard Model}

\subsubsection{The Higgs boson}

\begin{figure}
\centering
\includegraphics[bb = 72 580 188 720]{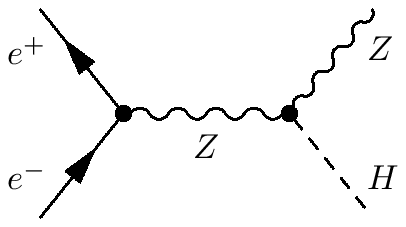}
\hspace{5mm}
\includegraphics{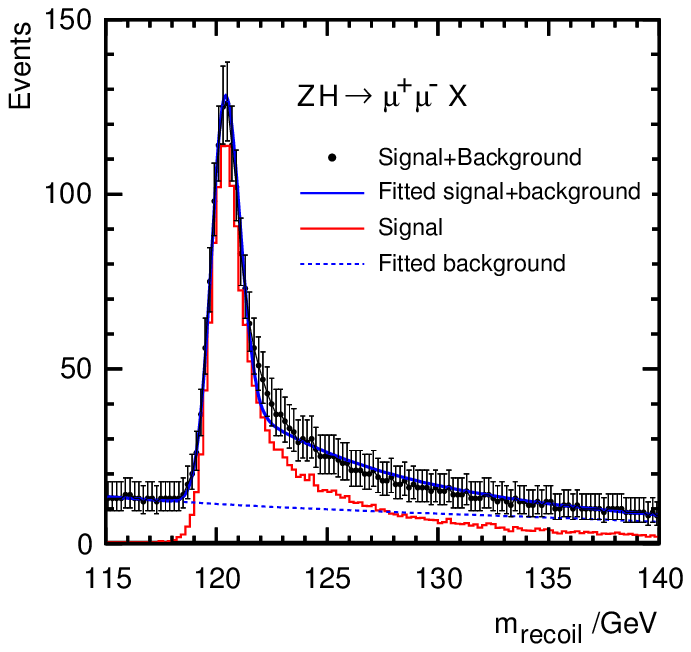}
\caption{\label{fig-higgsstrahlung}Left: Higgsstrahlung Feynmann diagram; right: recoil-mass distribution of muon pairs from the $Z$ decay at 250 GeV ILC (figure taken from Ref. \protect\cite{ILDLoI}, see also \protect\cite{Li09}).}
\end{figure}

The program of precise measurements in the Higgs sector is an excellent illustration of the capacity of the linear collider to advance our understanding of particle physics. The entry point into this field is the Higgsstrahlung process, in which a $Z$ boson is created in the annihilation of the initial electron-positron pair and emits a Higgs boson in the final state (Fig. \ref{fig-higgsstrahlung}, left). Experimental identification of the Higgsstrahlung is achieved by selecting lepton pairs with invariant mass consistent with the $Z$ mass. The distribution of the recoil mass, calculated under the assumption that all events occur at the nominal CM energy, features a clear peak at the Higgs mass, and a high-energy tail due to the luminosity spectrum (Fig. \ref{fig-higgsstrahlung}, right). In the analysis of the $Z\rightarrow \mu^+\mu^-$ decay, the absolute value of $g^2_{HZZ}$ is determined from the number of events in the peak with a precision of 2.5\% at ILC \cite{ILCTDR2} and 4.2\% at CLIC \cite{CLIC_snowmass13}. The Higgs mass is determined from the position of the peak with a statistical precision of 40 MeV at the 250 GeV ILC \cite{ILCTDR2} and 120 MeV at the 350 GeV CLIC \cite{CLIC_snowmass13}. If the analysis of the $Z\rightarrow e^+e^-$ decay is combined, the precision reaches 32 MeV at ILC \cite{ILCTDR2}.

\begin{figure}
\centering
\includegraphics{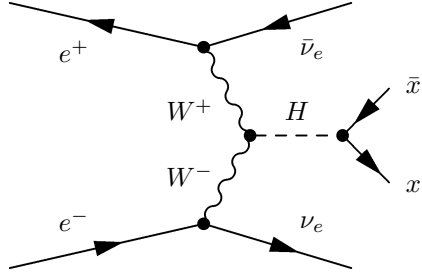}
\caption{\label{fig-WWfus}Feynmann diagram for Higgs production by $WW$ fusion, and subsequent decay to a particle-antiparticle pair.}
\end{figure}

At CM energies of 350 GeV and higher, Higgs production by $WW$ fusion allows for the measurement of Higgs couplings via the branching ratios (BR) for the Higgs decay to a pair of fermions or gauge bosons (Fig. \ref{fig-WWfus}). 

Using the value of $g^2_{HZZ}$ obtained in the Higgsstrahlung measurement, the Higgs coupling to $W$ is obtained from the relationship,
\begin{equation}
\centering
\frac{\sigma(e^+e^- \rightarrow ZH) \times BR(H \rightarrow x\bar{x})}
     {\sigma(e^+e^- \rightarrow \nu_e \bar{\nu_e} H) \times BR(H \rightarrow x\bar{x})}
\propto
\left( \frac{g_{HZZ}}{g_{HWW}} \right)^2 ,
\end{equation}
where the best statistical precision is reached in the case when $x$ stands for the $b$ quark. Higgs total decay width, $\Gamma_H$, can be obtained from either the $H \rightarrow WW^*$ or the $H \rightarrow ZZ^*$ decay,

\begin{equation}
\centering
\sigma(e^+e^- \rightarrow \nu_e \bar{\nu_e} H) \times BR(H \rightarrow WW^*)
\propto
\frac{g^4_{HWW}}{\Gamma_H}.
\end{equation}
Finally, $\Gamma_H$ can be used to determine the absolute value of all other measured couplings.

\begin{figure}
\centering
\includegraphics{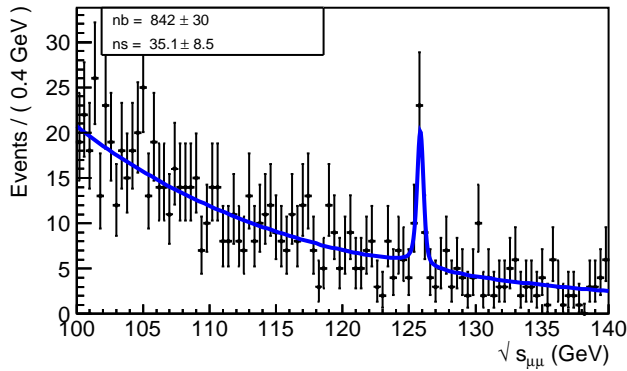}
\caption{\label{fig-mumufit}Fit of the model of the dimuon invariant mass distribution of the signal on top of the background remaining after an MVA selection at 1.4 TeV CLIC.}
\end{figure}

At 1 TeV or above, the cross section for the $WW^*$ fusion process is sufficiently high to allow for the measurement of rare Higgs decays such as the decay to a pair of muons, for which the BR is calculated to be $2.14 \times 10^{-4}$ \cite{LHC_higgs-xs}. In such measurements, after subtraction of background by selection cuts or multivariate analysis (MVA), the shape of the dimuon invariant mass distribution of the signal on top of the remaining background is fitted to the data (Fig. \ref{fig-mumufit}). The statistical precision of BR($h \rightarrow \mu\mu$) is 32\% at the 1 TeV ILC \cite{ILCTDR2}, 29\% at the 1.4 TeV CLIC \cite{CLIC_snowmass13}, and 16\% at the 3 TeV CLIC \cite{CLIC_snowmass13, Gre11}. An overview of achievable uncertainties in various Higgs measurements can be found in the ILC Technical Design Report \cite{ILCTDR2}, as well as in the CLIC Snowmass paper \cite{CLIC_snowmass13}.

\subsubsection{Top pair threshold scan}

\begin{figure}
\centering
\includegraphics{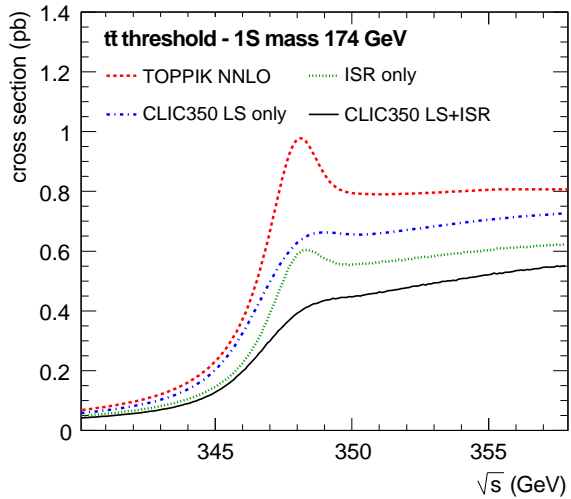}
\caption{\label{fig-topscan}Influence of the luminosity spectrum on the measured cross-section curve in the top-pair threshold scan (taken from Ref. \protect\cite{Seidel13}).}
\end{figure}

The top pair threshold scan is a very precise method of experimental determination of the top quark mass. The effective cross section for top pair production is measured in several energy points near the threshold, with $\sim 10 \text{fb}^{-1}$ of dedicated beam time per point. The position of the rising edge of the measured cross-section curve is sensitive to the top-quark mass. The precise value of the mass is extracted by fitting the theoretical calculation of the cross-section curve to the data. The luminosity spectrum, as well as the initial-state radiation (ISR) distribution have to be taken into account in the calculation, as can be seen in Fig. \ref{fig-topscan}. The statistical uncertainty of the top mass obtained in this way at either ILC or CLIC is 34 MeV. The overall uncertainty is, however, limited by the uncertainties of the theoretical calculation to about 100 MeV \cite{Seidel13}.

\subsection{Searches for new physics}
\label{sec-newphys}

Open questions in particle physics itself, as well as observations in other fields, such as cosmology, indicate that the SM does not cover all known phenomena. Such questions include the gauge hierarchy problem of the SM, the nature of elementary constituents of Dark Matter in the universe, or the source of CP violation in the evolution of the universe. Existing theoretical extensions of the SM that seek to address these questions drive the program of searches for new physics at future collider experiments.

\subsubsection{New gauge boson}

A common consequence of many extensions of the SM, in which the SM gauge group is embedded into a larger mathematical structure with additional U(1) symmetry groups, is the existence of one or more new, heavy and electrically neutral gauge bosons, denoted $Z'$. If the mass of $Z'$ is within the kinematical reach of the collider, it is observable as a resonance in the $e^+e^- \rightarrow f\bar{f}$ channel. However, even if the $Z'$ mass is higher than the CM energy of the collider, its existence can be observed via loop corrections of the $e^+e^- \rightarrow f\bar{f}$ cross section. The mass scale at which $Z'$ is detectable by such effects depends on the precision of the $e^+e^- \rightarrow f\bar{f}$ measurement. Depending on the model, the sensitivity of the 500 GeV ILC to the $Z'$ boson reaches between 4 and 10 TeV in terms of the 95\% CL for exclusion \cite{ILCTDR2}. The reach of the 1 TeV ILC is almost twice as high. At 3 TeV CLIC, depending on the couplings of $Z'$ to fermions, $5\sigma$ discovery of the $Z'$ boson will be possible for $m_{Z'}$ between 8 and 50 TeV, using the measured cross sections and forward-backward asymmetries \cite{CLIC_snowmass13}.

\subsubsection{Supersymmetry}

Supersymmetric theories postulate symmetry between bosons and fermions at a TeV scale. They offer a natural candidate for dark matter, as well as a possibility of unification of forces at high energies. The potential for discovery of the supersymmetric partners of the SM leptons is higher at CLIC than at the 14 TeV LHC (Tab. \ref{tab-reach}, \cite{CLIC_snowmass13}). Depending on the supersymmetric model, the production threshold for the lightest sparticles, which have not yet been excluded by the LHC, lies just below 1 TeV, requiring linear colliders of 1 TeV or more for their discovery. If supersymmetric particles are discovered, linear colliders offer unique opportunity to measure their masses and couplings, and thus test the existing theories. 

\subsubsection{Discovery reach}

\begin{table}
\centering
\includegraphics{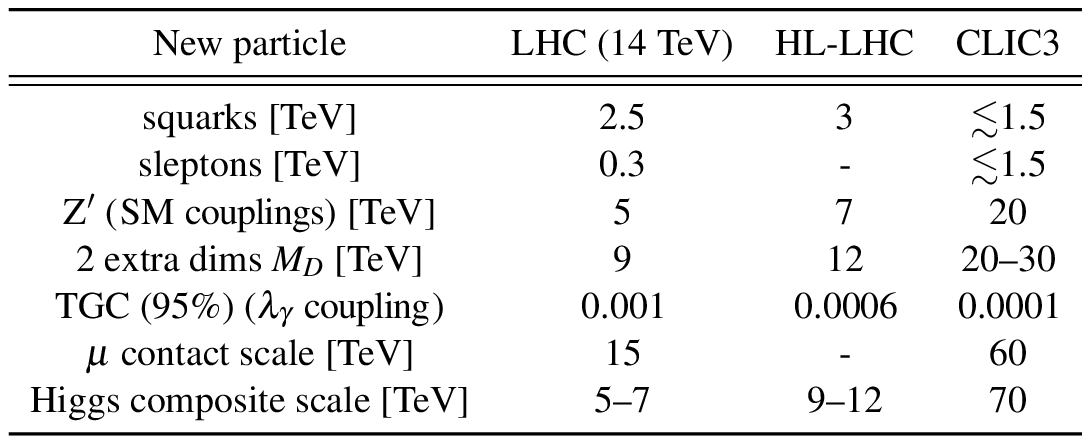}
\caption{\label{tab-reach}Discovery reach in various theory models for different colliders. LHC at 14 TeV refers to integrated luminosity of 100 $fb^{-1}$, HL-LHC 1 $ab^{-1}$, and the 3 TeV CLIC up to 2 $ab^{-1}$. Taken from Ref. \protect\cite{CLIC_snowmass13}.}
\end{table}

A brief overview of the discovery reach of the 3 TeV CLIC in comparison with the LHC and the HL-LHC is given in Tab. \ref{tab-reach} from Ref. \cite{CLIC_snowmass13}. Beside the discovery potential for the supersymmetric particles, and the $Z'$ boson, energy scale for theories with extra spatial dimensions is listed, the sensitivity level for anomalous triple coupling of the gauge bosons (TGC), the $\mu$ contact interaction scale, as well as the composite Higgs boson mass scale \cite{CLICCDR12PD}.

\section{Conclusions}

In this lecture, basic motivation for building a next-generation linear collider was given, together with the accelerator and detector concepts. The physics program was outlined in its main aspects, including the SM studies and the search for new physics, and several higlights from the benchmark studies were given. In these benchmark studies, based on detailed and realistic simulations, the precision capabilities of linear colliders were confirmed and significant discovery potential has been demonstrated for the searches for new physics. Linear collider is a promising tool, complementing the LHC in essential ways to test the SM and to search for new physics. 

The linear collider study is a broad field for R\&D in accelerator technology and in detector hardware, as well as for physics analysis work. Once built, the physics program of the linear collider unfolds in energy stages, and spans over 20 years of research work with potentially ground-breaking physics opportunities at each stage.

\providecommand{\noopsort}[1]{}\providecommand{\singleletter}[1]{#1}%
\providecommand{\href}[2]{#2}\begingroup\raggedright\endgroup

\end{document}